# Development and Validation of Patient-Specific Monte Carlo Dosimetry for Synchrotron Breast Phase-Contrast CT

Amir Entezam[1,2], Ashkan Pakzad[1], Christopher J. Hall[2], Anton Maksimenko[2], Matthew John Cameron[2], Adam Round[2], Mojtaba Hoseini-Ghahfarokhi3, Seyedamir T. Taba4, Yakov I. Nesterets5, Daniel Häusermann[2], Magdalena Bazalova-Carter6, Patrick C. Brennan[3], Timur E. Gureyev[1], Harry M. Quiney[1]

[1] School of Physics, The University of Melbourne, Parkville, VIC 3010, Australia

[2] Australian Synchrotron, ANSTO, Clayton, VIC 3168, Australia

3 Department of Radiation Physics, The University of Texas, MD Anderson Cancer Center, Houston, Texas 77030, USA

4 Faculty of Health Sciences, The University of Sydney, Lidcombe, NSW 2141, Australia

5 Commonwealth Scientific and Industrial Research Organisation, Clayton, VIC 3168, Australia

6 Department of Physics and Astronomy, University of Victoria, Victoria, British Columbia, V8P 5C2, Canada

**Abstract**

*Objective*: To develop and validate a patient-specific Monte Carlo (MC) dosimetry framework for propagation-based phase-contrast breast computed tomography (BCT) at the Imaging and Medical Beamline (IMBL) of the ANSTO Australian Synchrotron, enabling accurate estimation of mean glandular dose (MGD) for clinical purposes.

*Background*: BCT offers three-dimensional imaging without breast compression, improving patient comfort and enhancing visualization of internal breast structures, which is important for accurate cancer detection. Propagation-based phase-contrast BCT provides superior soft-tissue contrast at comparable or lower doses than conventional absorption-based BCT. Accurate radiation dosimetry remains essential for ensuring patient safety and optimizing imaging protocols. Most MC-based studies of MGD have not used realistic patient-derived phantoms, thereby neglecting anatomical variability that can significantly affect dose distribution, while the limited studies incorporating patient-specific phantoms rely on approaches that lack a comprehensive, reproducible, and transferable framework for reliable dosimetry.

*Methods*: To address this limitation, a voxel-based MC dosimetry framework was developed using the EGSnrc MC code to compute MGD in realistic anthropomorphic digital breast phantoms derived from synchrotron BCT images. The IMBL beam characteristics were defined as the source parameters in the simulations. Additionally, corresponding homogeneous phantoms were generated, and the air kerma ($K_{air}$) to MGD conversion coefficients, expressed as normalized glandular dose coefficients (DgN), were calculated and compared with those from heterogeneous phantoms. MC simulations were conducted for varying breast heights and skin thicknesses over an energy range of 28–38 keV.

*Results*: Breast anatomy and X-ray beam energy substantially affected the MGD. Increased glandular density led to lower MGD values, whereas larger breast volume resulted in higher MGD values. A 2 mm increase in skin thickness produced approximately a 10% decrease in MGD. The comparison between heterogeneous and homogeneous phantoms demonstrated measurable variations in DgN, highlighting the importance of incorporating anatomical heterogeneity in dosimetric modelling.

*Conclusion*: The developed voxel-based MC framework provides a robust and anatomically accurate foundation for patient-specific dosimetry in synchrotron phase-contrast BCT. It enables precise MGD estimation and supports the safe, effective implementation of clinical breast imaging.

**1. Background**

Breast cancer is the most frequently diagnosed cancer among women worldwide and remains one of the most serious global health concerns [1]. In Australia, breast cancer was the leading cause of cancer-related death among females aged 45–64 in 2023 [2]. Although screening programs have greatly improved early detection, diagnostic challenges remain, particularly in women with dense breast tissue or lesions that are difficult to visualize [3,4]. Currently, the primary imaging modalities for breast cancer screening are two-dimensional (2D) Digital Mammography (DM) and three-dimensional (3D) Digital Breast Tomosynthesis (DBT) [5]. DM produces high-resolution 2D projection images of the compressed breast, while DBT acquires projections over a limited angular range to reconstruct pseudo-3D slices, thereby improving lesion visibility [5–7].

To address the inherent limitations of 2D projection techniques, particularly tissue superposition, fully 3D imaging methods such as X-ray computed tomography (CT) have been developed [8,9]. These approaches provide cross-sectional images with enhanced anatomical accuracy and improved contrast [10]. BCT is a specialized version of CT designed specifically for breast imaging [11,12]. One widely used implementation is cone-beam BCT (CB-BCT) [13].

BCT acquires high-resolution volumetric data by rotating either the X-ray source or the patient through 360°, capturing multiple projections from different angles [14]. Unlike DM and DBT, BCT eliminates breast compression [9], improving patient comfort and removing tissue overlaps, which enhances visualization of dense and anatomically complex structures.

The clinical utility of BCT has been increasingly recognized, with regulatory approvals granted in both the United States [15] and Europe [16].

Phase-contrast CT (PCT) is an advanced imaging technique that significantly enhances soft-tissue contrast. Unlike DM, DBT, and BCT, which rely solely on X-ray absorption differences, PCT captures additional image contrast arising from phase shifts caused by variations in tissue refractive indices associated with differences in composition and density [17]. This additional phase information enhances image quality and soft tissue contrast without the need to increase the radiation dose [18]. Among phase-contrast techniques, propagation-based PCT (PB-PCT) is attractive because it relies only on free-space propagation without additional X-ray optical components [19,20]. It relies on a phase retrieval process to reconstruct phase information from the measured intensity data, making it an appropriate technique for high-resolution and low-dose imaging, especially in breast and lung imaging applications [21].

In this study, a PB-PCT technique was employed to establish a dosimetry workflow at the IMBL beamline of the ANSTO Australian Synchrotron for breast imaging. The IMBL beamline uses a superconducting multipole wiggler capable of delivering high-intensity, high-energy, quasi-parallel X-ray beams [22]. Such beams are ideally suited for propagation-based phase-contrast imaging, enabling stable, high-quality data acquisition under well-controlled imaging conditions [22].

Radiation exposure in breast imaging carries a risk of radiation-induced carcinogenesis, as breast tissue is highly sensitive to ionizing radiation [23]. Accurate dose estimation is therefore critical. The standard dose metric for breast imaging is the MGD, expressed in milligray (mGy). MGD represents the average energy absorbed per unit mass of glandular tissue, the most radiosensitive tissue of the breast, and serves as the primary indicator of breast radiation risk. The U.S. Food and Drug Administration (FDA) has defined a regulatory dose limit of 3 mGy per view for both DM and DBT to reduce radiation exposure risks to patients [24]. MGD is commonly estimated as the product of the incident $K_{air}$ and the DgN:

$$\mathrm{MGD} = \mathrm{DgN} \times \mathrm{K_{air}} \quad \text{(Equation 1)}$$

where both MGD and $K_{air}$ are expressed in mGy, and DgN (dimensionless) accounts for factors such as breast size, composition, X-ray energy, and imaging modality. Established procedures and tabulated DgN values exist for DM, DBT, and BCT [25–30]. Only the incident $K_{air}$ is measured as part of the procedure.

Because direct *in vivo* measurements of absorbed dose and glandular tissue mass are not feasible, MGD is typically estimated via MC simulations based on computational breast models. In these simulations, DgN values are calculated using MC algorithms such as MCNP [31], EGSnrc [32], and, most notably, Geant4 [33]. These DgN values must be adapted to account for specific breast and X-ray beam characteristics and the imaging modality used [34].

MC simulations for MGD estimation have so far mostly used simplified breast models that assume a homogeneous mixture of glandular and adipose tissues, and with or without a skin-equivalent outer layer [35–39]. While these models help standardize dose calculation, they fail to capture the anatomical complexity seen in clinical breast imaging. In reality, accurate MGD estimation is strongly influenced by multiple breast-specific characteristics. One key factor is breast density, which refers to the relative proportion of glandular and adipose tissue. Before applying DgN values, it is essential to determine the density of each individual breast. Earlier studies often assumed a uniform glandularity of 50% across different breast sizes, based on population-level mammography data [40]. However, recent clinical investigations show that this assumption may not accurately represent the average breast densities [41].

Simplified homogeneous breast models also do not account for the actual spatial distribution of glandular tissue, which can influence MGD values for a given imaging technique and breast shape. Glandular tissue is typically distributed through the breast non-uniformly, often concentrated at the center of the breast and partially shielded by adipose tissue [35]. This anatomical feature affects the internal dose distribution and can lead to systematic and substantial misestimation of MGD, an issue highlighted in several studies. For instance, homogeneous models in mammography have been shown to result in MGD deviations of up to 50% [34,42]. Another study reported that assuming uniform glandular distribution could result in MGD conversion factor errors as high as 48% [43]. In BCT, the use of homogeneous models results in smaller errors in MGD estimation compared to DM and DBT. This is primarily due to the rotational scanning geometry, which reduces the dependence of absorbed dose on the spatial distribution of glandular tissue [44]. Studies have shown that the underestimation of MGD in CB-BCT when using homogeneous models is less pronounced, with errors of approximately 10% at 49 kVp, an operating voltage typical of systems such as those delivered by the X-ray scanners from Koning Corporation [40,45].

Several studies have shown that the DgN is highly sensitive to breast size and shape [36,37]. To address this, the American Association of Physicists in Medicine (AAPM) Task Group 195 (TG195) established guidelines for constructing breast phantoms and computational models in breast imaging. This widely adopted model, described by Sechopoulos *et al.* (2015), uses a cylindrical breast model with a 10 cm diameter as a reference geometry to standardize dosimetric assessments across imaging systems [38].

In practice, many CB-BCT studies employ semi-ellipsoid or cylindrical geometries to more closely approximate the pendant-geometry breast during imaging [37,40,46]. In these models, the minor axis corresponds to the radius of the semi-ellipsoidal breast at the chest wall, while the major axis corresponds to the distance from the chest wall to the nipple [40]. Clinical data report an average chest wall radius of about 7 cm and an average chest wall-to-nipple distance of approximately 10.5 cm [46,40]. Also, a semi-ellipsoidal model was introduced that incorporated breast length and evaluated DgN for monoenergetic beams, showing that DgN increases as breast diameter decreases [40]. Although these geometric refinements improve accuracy, most models still assume a uniform mix of glandular and adipose tissue composition.

Accurate modelling of skin is another important factor in reliable MGD estimation. Traditional dosimetric protocols, such as TG195, often assume a skin thickness of 2 mm, consisting of equal proportions of glandular and adipose tissue. However, Sechopoulos *et al.* (2020) report that the skin envelope often contains a higher proportion of either glandular or adipose tissue, rather than a balanced mixture [39]. This tissue-specific variation within the skin layer further affects MGD estimation [40]. Shi *et al.* (2013) used high-resolution BCT to measure breast skin thickness and reported a mean value of 1.44 ± 0.25 mm [47]. This is relatively consistent with the range of 0.5 mm to 2 mm reported in earlier literature [38]. To create a more realistic breast model, Dance *et al.* (2016) incorporated a 1.45 mm thick skin layer into their breast models for MGD estimation in the Koning BCT system [48]. Their MC simulations showed that reducing skin thickness from the standard 4 mm to a more anatomically accurate 1.5 mm can increase DgN values by up to 18%. This increase is especially pronounced in smaller breasts, such as those with a maximum of 5 cm radius.

These findings underscore the need for anatomically accurate, patient-specific breast models that reflect actual breast characteristics. When combined with MC simulations under clinically representative imaging conditions, these models can enable individualized and reliable MGD estimation. As Hammerstein *et al.* (1979) stated: "Detailed information will have to be obtained on the amount and distribution of gland tissue in many individual cases before energy absorbed in glandular tissue can be applied properly to the problem of individual risk" [49]. Several previous studies have used CT-derived realistic breast phantoms to estimate MGD in clinical polychromatic imaging systems, such as CB-BCT systems operating with W/Al spectra at approximately 49 kVp, including those developed by Koning [40,36]. However, no MC framework to date has been specifically developed or validated for MGD estimation in synchrotron-based monoenergetic PCT, particularly using realistic breast phantoms. Synchrotron X-ray sources exhibit fundamentally different beam characteristics compared to conventional tube-based systems, including quasi-parallel beam geometry, high photon flux, and strictly monochromatic energies. These features result in distinct dose deposition patterns and can deliver comparable or higher doses over much shorter acquisition times [22,50]. Consequently, accurate MGD estimation, accounting for both system-specific and patient-specific factors, is particularly critical for synchrotron-based breast imaging.

Although some studies have combined realistic breast phantoms with commercial software such as ImpactMC (CT Imaging GmbH, Erlangen, Germany) for dose estimation in uncompressed breast imaging, these tools are primarily optimized for tube-based CB-BCT [51,52]. As a result, they are not well suited to model essential aspects of synchrotron imaging, including monochromatic spectra and specialized beam geometries. Moreover, their dose-scoring methods are based on assumptions specific to conventional imaging modalities, further limiting their applicability to synchrotron-based PCT. In addition, the restricted availability of such commercial software, compared with widely accessible general-purpose MC codes, limits their practical use for MGD estimation in specialized synchrotron imaging research. Few studies, such as Franco *et al.* (2022), have combined realistic breast phantoms with advanced virtual clinical trial platforms like VICTRE to estimate MGD [53]. VICTRE offers a comprehensive framework that includes anatomical modelling, projection simulation, noise modelling, and image reconstruction. However, its complexity and high computational requirements make it less practical for focused dosimetric studies using digital phantoms, limiting its adoption by other research groups for independent MGD estimation.

Another limitation in previous studies is the minimal investigation of how skin thickness affects MGD when using realistic breast phantoms. Implementing an actual realistic skin layer into an existing voxel-based breast phantom is technically challenging, and assessing its effect across a wide range of thicknesses usually requires large datasets to achieve statistically meaningful results.

All previous studies reviewed in this work that used realistic breast phantoms for MGD estimation have relied almost exclusively on the Geant4 MC toolkit. While Geant4 provides flexibility and accuracy, its dominance has left other capable MC platforms underexplored in MGD calculations. In this work, the EGSnrc MC code was employed for breast dosimetry due to several advantages relevant to medical imaging applications. EGSnrc offers high flexibility for handling large CT-derived voxelized phantoms, enabling realistic patient-specific dose calculations, and is well known for its accurate modelling of low-energy photon transport, which is particularly important for kilovoltage breast imaging [32]. By demonstrating its applicability in this context, this study broadens the range of validated MC tools and provides an alternative computational framework for MGD assessment in breast imaging.

We present a MC framework for MGD estimation in synchrotron-based phase-contrast breast CT using realistic voxel-based breast phantoms and EGSnrc simulations. The proposed approach incorporates patient-specific breast anatomy, tailored to the BCT setup at the IMBL of the ANSTO Australian Synchrotron. This methodology enables accurate 3D dose mapping and supports reproducible, patient-specific dosimetry. It further allows systematic evaluation of the influence of anatomical heterogeneity and skin thickness on MGD estimation while preserving internal tissue structure. Although developed for the BCT setup at the IMBL, the framework can be readily adapted to other breast imaging systems or synchrotron-based medical imaging beamlines for precise dosimetry.

## 2. Materials and Methods

### 2.1 Voxel-based realistic breast phantoms

#### 2.1.1 Specimen preparation and imaging

Realistic digital anthropomorphic breast phantoms to be used in the MC calculations were developed from high-resolution CT imaging of human mastectomy specimens using PB-PCT at the IMBL. The imaging experiments were conducted under a Human Ethics Certificate of Approval from Monash University (Project ID 26399). Written consent was obtained from all patients to image their clinical specimens.

Five freshly excised human mastectomy specimens, labeled Samples 1–5, were obtained post-surgery for this study. Each specimen was positioned without compression inside a plastic cylindrical holder with a fixed diameter of 11.4 cm, which is approximately consistent with the average breast diameters reported by Mettivier *et al.* (2016) [36] and Sarno *et al.* (2018) [45]. The axis of the cylinder was perpendicular to the beam direction and parallel to the rotation axis. When mounted in the holder, the specimen height, defined as the distance from the chest wall to the nipple along the cylinder axis, was 6.5 cm for Samples 1 and 2, and 5 cm, 7 cm, and 9 cm for Samples 3–5, respectively, reflecting inter-subject variations in breast volume. All specimens were oriented with the nipple directed superiorly. The glandular fractions were 10% and 25% for Samples 1 and 2, and 40%, 30%, and 25% for Samples 3–5, respectively, as summarized in Table 1.

| Sample | Holder Diameter (cm) | Specimen Height (cm) | Glandular Fraction (%) |
|---|---|---|---|
| 1 | 11.4 | 6.5 | 10 |
| 2 | 11.4 | 6.5 | 25 |
| 3 | 11.4 | 5.0 | 40 |
| 4 | 11.4 | 7.0 | 30 |
| 5 | 11.4 | 9.0 | 25 |

***Table 1. Characteristics of the specimens used in this study***

During imaging, the specimens were rotated 180° around the vertical axis to obtain complete tomographic datasets. The same imaging geometry was reproduced in MC simulations to ensure anatomical and dosimetric consistency. A total of 4800 angular projections were uniformly acquired over 180° at beam energies 28–38 keV, delivering an MGD of 24 mGy. This value was estimated using an average DgN of 0.5, representing a mean across the studied conditions, together with an incident air kerma of 48 mGy. The actual MGD may vary slightly between samples due to differences in breast composition and geometry. However, precise dose accuracy is not critical at this stage. The selected dose level only provides a high contrast-to-noise ratio, enabling accurate tissue segmentation and reliable phantom generation. Images were recorded using a flat-panel energy-integrating detector (Teledyne DALSA Xineos 3030HR) with a 99 µm pixel pitch [22]. Projection data were pre-processed prior to 3D reconstruction to correct for dark-current and flat-field variations, inter-pixel gaps, and defective pixels. Image reconstruction was performed using filtered back projection (FBP) combined with the TIE-Hom phase-retrieval algorithm [22]. Detailed descriptions of the sample imaging, reconstruction process, and image quality assessment are provided in Ref. [54].

The resulting volumetric datasets preserved the native anatomical heterogeneity of uncompressed human breast specimens, accurately representing glandular, adipose, and skin structures. These datasets formed the basis for constructing realistic voxelized digital breast phantoms for subsequent MC-based dosimetric evaluations.

**2.1.2 EGSnrc-compatible voxel phantoms**

Voxelized breast phantoms were generated from the reconstructed volumetric datasets for subsequent MC EGSnrc code system simulations using the custom MATLAB scripts (R2017b, MathWorks, Natick, MA, USA). The custom MATLAB code requires input images in DICOM format. These images, encoded in Hounsfield Units (HU), were then converted into material type and density maps using a user-defined calibration ramp.

In this work, PCT reconstructions produced β-maps that are directly proportional to the local linear attenuation coefficient and describe the spatial distribution of the imaginary component of the complex refractive index within breast tissue. The complex refractive index $n$ is expressed as

$$\mathrm{n} = 1 - \delta + \mathrm{i}\beta \qquad \text{(Equation 2)}$$

where $\delta$ is the refractive index decrement (real part), responsible for phase shifts, and β is the absorption index (imaginary part), which characterizes X-ray attenuation at a given energy. Consequently, the β value at each voxel reflects the intrinsic absorption properties and composition of the tissue. The β values were converted to HU-equivalent values using Equation (3). The conversion is based on the relation between the β and the linear attenuation coefficient, $\mu = 2k\beta$, where $k$ is the X-ray wavenumber. Substituting this relation into the conventional CT definition of HU yields

$$\mathrm{HU} = 1000\,\frac{(2\mathrm{k}\beta - \mu_{\mathrm{water}})}{\mu_{\mathrm{water}}}. \qquad \text{(Equation 3)}$$

Where $\mu_{water}$ is the linear attenuation coefficient of water at the corresponding photon energy (28–38 keV), obtained from the NIST database [55].

The processed data were saved as coronal DICOM slices and imported into MATLAB scripts, where the HU-based DICOM datasets were converted into EGSnrc-compatible voxelized phantoms and written to egsphant files (the EGSnrc phantom file format). Based on its HU value, each voxel was assigned a material type and corresponding mass density. The material types included AIR521ICRU and ADIPOSE521ICRU, which are available in the 521icru PEGS4 data file, as well as glandular and skin tissues. The latter were manually defined using elemental compositions and mass densities obtained from NIST [56] and subsequently added to the material library via the EGS-GUI interface. The HU-to-material conversion was implemented using ranges informed by reported CT attenuation values of breast tissues [57]. The calibration ramp was defined as follows: -1000 to -700 HU for air, -699 to -100 HU for adipose tissue, -99 to +80 HU for glandular tissue, and +81 to +200 HU for skin. These extended ranges are consistent with prior CT-based breast tissue characterization studies and allow for variations arising from imaging conditions, reconstruction, and patient-specific anatomy [58,59]. Also, to evaluate the sensitivity of the results to the selected HU thresholds, a perturbation analysis was performed in which all HU boundaries were changed by ±40 HU. The resulting changes in calculated dose metrics were negligible (differences < 1%), indicating that the simulation outcomes are not sensitive to reasonable variations in HU-to-material assignment. The tissue definitions are summarized in Table 2.

| **Tissue Type** | **Material Name** | **Density (g/cm³)** | **Elemental Composition (% by weight)** | **Source** |
|---|---|---|---|---|
| Air | AIR521ICRU | 0.001204 | N: 75.5%, O: 23.2%, Ar: 1.3% | ICRU37 |
| Adipose Tissue | Adipose521ICRU | 0.95 | H: 11.4%, C: 59.8%, N: 0.7%, O: 27.8%, Na: 0.1%, P: 0.2% | ICRU44 |
| Glandular Tissue | Glandular | 1.04 | H: 10.5%, C: 63.0%, N: 2.0%, O: 23.5%, Na: 0.2%, P: 0.8% | NIST |
| Skin | Skin | 1.09 | H: 10.2%, C: 61.5%, N: 1.9%, O: 26.0%, Na: 0.2%, S: 0.2% | NIST |

***Table 2. Material assignments details used for tissue segmentation.***

Each voxel in the resulting egsphant files had a voxel size of 0.16 × 0.16 × 1 mm³ in the x, y, and z directions, respectively. The coronal slice dimensions were 713 × 713 voxels in the x–y plane, corresponding to the 11.4 cm diameter of the sample holder. The number of slices along the z-axis varied according to each sample size, resulting in 65, 65, 50, 70, and 90 slices for Samples 1–5, respectively. These values corresponded to holder heights of 6.5, 6.5, 5.0, 7.0, and 9.0 cm, ensuring that the full specimen volume was accurately represented. This voxel configuration balanced anatomical detail with computational efficiency in MC simulations [60,61]. Four types of phantoms were generated for comparative analysis:

1. *Heterogeneous phantoms:* Each voxel retained its original mass density value and the corresponding HU range, preserving the natural spatial variations in tissue density. The phantom closely mimicked the actual breast, accurately reflecting the anatomical distribution of glandular, adipose, and skin tissues. This ensured that the density, realistic spatial arrangement, and structural characteristics of the breast were faithfully represented.
2. *Material-homogeneous phantoms:* Each tissue type (adipose, glandular, skin) was assigned a single mean density value, calculated from the corresponding voxels in the heterogeneous phantom. While the characteristic location of each tissue type was maintained, all voxels within a given tissue were set to this single mean value, thereby removing intra-tissue density variations.
3. *Homogeneous phantoms:* In these phantoms, adipose and glandular voxels were assigned the single mean mass density value derived for the material-homogeneous phantom. Unlike the material-homogeneous phantom, these voxels were uniformly distributed across the entire breast phantom volume, representing a fully mixed composition. This uniform distribution did not reflect realistic anatomical placement of gland and adipose tissues but aligned with the homogeneous phantom fabrication approaches reported in several previous studies on MGD estimation [35].
4. *Independent MATLAB-based homogeneous phantoms:* These phantoms correspond to the voxelized homogeneous phantoms described above (Phantom 3) but are generated independently, without relying on patient-specific datasets. To create these phantoms, only a few key parameters are required: the mass densities of glandular and adipose tissues (obtained from NIST), overall breast characteristics (including diameter and height), and the glandular fraction of the breast. Using this information, the phantoms are computationally constructed with adipose and glandular voxels uniformly distributed throughout the breast volume, ensuring that the total glandularity and geometric dimensions match those of the corresponding homogeneous phantom. This approach enables independent generation of homogeneous phantoms without the need for realistic patient-derived phantoms or CT datasets, while maintaining consistent material properties and anatomical proportions with the original homogeneous models. Such phantoms provide a flexible framework for simulating MGD based on homogenous models across different breast sizes and compositions.

#### 2.1.3 Skin layer segmentation and generation

Mastectomy specimens often retain only partial skin coverage due to skin-sparing surgical techniques, whereas *in vivo* breasts are fully enclosed by skin. To enable accurate MC dose simulations that account for the skin, realistic synthetic skin envelopes were reconstructed around each specimen. Segmentation was performed to generate non-overlapping masks for the skin tissue. First, the breast was segmented from the β-map using intensity thresholding, connected-component analysis, and morphological refinement. The residual native skin along the superior surface was isolated, and voxel-wise β-statistics (mean and standard deviation) were extracted to guide the creation of synthetic skin textures that reflect physiological variability.

To model skin envelopes with thicknesses of 1, 2, and 3 mm, the native skin was digitally removed from both the CT image and the corresponding segmentation masks. This was achieved by first segmenting the skin and then assigning it a background value of 0 in the CT image, effectively removing the skin digitally. A new skin layer was then generated by dilating the breast surface using a spherical structuring element corresponding to the desired thickness, ensuring a uniform and anatomically consistent skin envelope for each model. Realistic skin texture was introduced by generating a random field sampled from a normal distribution, followed by Gaussian smoothing using an isotropic kernel with a standard deviation of 1 voxel, ensuring spatial continuity while preserving statistically realistic variation. The resulting voxel values were then rescaled to match the β-distribution of the native skin in terms of mean and standard deviation, producing a statistically and morphologically realistic synthetic layer.

All processing was implemented in Python 3.9.10 using NumPy 1.26.1, SciPy 1.11.3, and scikit-image 0.24.0, with 3D visualization performed in 3D Slicer 5.6.2.

**2.2 MC simulation of the imaging system**

MC modelling of the synchrotron-based BCT imaging beam at the IMBL was performed using BEAMnrc code (version 10.3) [32]. The objective was to reproduce the effective X-ray beam conditions at the sample position during PB-PCT experiments. To achieve this, a phase-space file, which is an EGSnrc scoring feature storing particle properties (including energy, position, and direction), was generated to represent the incident photon field at the breast entrance plane. The phase space file was used as input in further simulations by DOSXYZnrc.

At the Australian Synchrotron IMBL, the X-ray beam is generated by a superconducting wiggler source located approximately 136m upstream of the experimental hutch used for this imaging [22]. However, explicit modelling of the entire beamline, including filtration, optical components, monochromation, beam expansion, and collimation, would introduce substantial geometric and computational complexity without materially affecting the beam characteristics at the sample plane relevant for dosimetry, as confirmed by experimental verification of the modelled beam properties presented in Section 2.3. Instead, the MC source was defined to accurately represent the final collimated beam at the sample position. Experimentally, the beam was first expanded to approximately 20 cm (horizontal) × 8 cm (vertical) using the IMBL beam-expansion system and subsequently shaped by precision slits to produce a nearly uniform rectangular field at the sample plane [22]. Breast specimens were positioned 2.7 m downstream of the final collimation stage on a rotating CT platform and imaged over 180°. At this location, the beam exhibited high spatial uniformity across the field-of-view. Accordingly, in the MC simulations, the beam was defined directly as a monochromatic and spatially uniform rectangular source with a size matching the collimated field (20 cm × 8 cm) and photon energies of 28, 30, 32, 34, 36, and 38 keV, corresponding to the synchrotron operating settings used experimentally. This approach ensured that the simulated source represents the experimentally delivered, fully collimated rectangular beam while avoiding unnecessary modelling of upstream beamline components that do not influence the absorbed dose distribution within the breast.

In BEAMnrc, the FLATFILT component module was used to generate a monoenergetic rectangular source matching the experimental beam dimensions (20 cm × 8 cm). A corresponding phase-space file was then generated for the modelled beam, with the scoring plane positioned 2.7 m downstream of the source to reproduce the experimentally collimated incident photon field at the entrance surface of the sample. A total of $5\times10^9$ primary photon histories were simulated to achieve adequate statistical precision and minimize MC uncertainties. For each particle crossing the scoring plane, its energy, spatial coordinates, and directional cosines were recorded. The resulting phase-space file therefore provided a complete description of the experimental incident beam prior to interaction with the specimen. This file can be reused as input for downstream MC dose calculation without re-simulating the upstream beamline.

Photon transport physics in BEAMnrc included all relevant low-energy photon-matter interactions appropriate for the investigated energy range, namely photoelectric absorption, coherent (Rayleigh) scattering, incoherent (Compton) scattering, and atomic relaxation with characteristic X-ray emission. Electron transport, including secondary electron production and energy loss, was also enabled to ensure accurate modelling of energy deposition in the breast tissues. Bremsstrahlung production from secondary electrons was included, although its contribution was negligible at these photon energies. The global transport parameters were set to ensure accurate particle tracking and dose calculation. The electron cut-off energy (ECUT) and electron production threshold (AE) were set to 0.521 MeV total energy, which corresponds to a 10 keV kinetic energy above the electron rest mass (0.511 MeV). The photon cut-off energy (PCUT) and photon production threshold (AP) were set to 0.01 MeV, consistent with standard EGSnrc settings for low-energy photon transport. These settings ensured that all relevant secondary particles were transported and that energy deposition within the breast model was accurately captured. Overall, the phase-space generation produced a statistically robust, spatially uniform, and monochromatic beam model, providing a reproducible input for subsequent MC dose simulations.

### 2.3 Validation of the MC model

To verify the accuracy of the phase-space generation and to validate the MC model of the imaging system, experimentally measured percentage depth dose (PDD) curves were acquired and compared with corresponding PDD curves calculated using MC simulations based on the generated phase-space file. Agreement between measured and simulated depth-dose distributions provides confidence that the incident beam model accurately represents the experimental imaging beam.

#### 2.3.1 Experimental PDD measurements

PDD measurements were performed using the IMBL water tank phantom, the specifications of which are described in [62]. The water tank was constructed from 4 mm thick Perspex polymethyl methacrylate (PMMA). The internal dimensions of the phantom are 158 × 137 × 110 $mm^3$. It was filled with Reverse-Osmosis water to approximate the radiological properties of soft tissue for depth-dose measurements in the diagnostic energy range [62].

Dose measurements were carried out using a PTW Farmer (TN: 30013) ionization chamber (PTW-Freiburg, Germany) [62]. The water tank was positioned at the center of the rectangular imaging beam to ensure accurate alignment of the detector with the beam axis. The ionization chamber was mounted on a motorized translation stage, enabling precise positioning along the depth direction of the water phantom. The chamber's sensitive volume was initially positioned at the entrance wall of the water tank to measure dose at the minimum depth. The detector was then translated along the depth axis in 5 mm increments, with measurements recorded at each depth. At each position, five repeated measurements were performed, and the mean value was reported to ensure improved statistical reliability. The collected ionization readings were converted to absorbed dose in water and corrected for ion recombination, temperature, and pressure, following the dosimetry protocol described in Ref. [63], yielding the experimental PDD curve.

#### 2.3.2 MC simulation of PDD curves

The corresponding PDD curves were calculated using the DOSXYZnrc code, part of the EGSnrc MC package [64]. The phase-space file generated as described in Section 2.2 was used as the incident source and was applied at normal (0°) incidence to the water tank, replicating the experimental irradiation geometry. The experimental water tank was modelled as a voxelized phantom in DOSXYZnrc, matching the physical dimensions and composition of the experimental setup. Voxel materials were defined as PMMA521ICRU for the PMMA walls, WATER521ICRU for the water volume, and ICRUAIR521 for the surrounding air, corresponding to the material identifiers used in the 521icru material library. The beam center point was aligned with the central axis of the phantom. Energy deposition was scored in each voxel, and the resulting three-dimensional dose distribution was stored in the standard .3ddose format. Using the DOSXYZnrc STATDOSE interface, a one-dimensional depth-dose curve was extracted along the central beam axis to generate the simulated PDD curve for direct comparison with the measured data. MC simulations were performed on an Intel Core Ultra 7 165H processor (16 cores, 3.8 GHz) and 16 GB RAM. Photon transport through the voxelized geometry was simulated using $5 \times 10^9$ photon histories. Statistical uncertainties were extracted from the standard DOSXYZnrc output and are reported as one standard deviation (1σ), remaining below 1% in the high-dose region, following recommended reporting practices (AAPM TG-268 [65]). The computation time was on the order of 5−7 hours per simulation, depending on the configuration.

#### 2.3.3 Gamma index comparison

Agreement between the measured and simulated PDD curves was quantified using gamma index analysis, a widely accepted method for dose distribution comparison [66]. The analysis employed criteria of ≤ 2 mm distance-to-agreement (DTA) and ≤ 2 % dose difference. Gamma values less than 1 were considered passing the test. The gamma analysis was performed using in-house MATLAB routines, with the experimentally measured PDD serving as the reference distribution.

### 2.4 MC dose calculation in breast phantoms

The voxelized breast phantoms described in Section 2.1 were imported into DOSXYZnrc for voxel-based dose scoring. The validated phase-space file described in Section 2.2, representing the monochromatic imaging beam, was used as the incident source for all dose calculations.

The geometric center of each breast phantom surface was identified using 3D Slicer and defined as the simulation isocenter. This isocenter was aligned with the midplane of the phase-space source to accurately replicate the experimental beam–phantom geometry. In DOSXYZnrc, the distance from the scoring plane to the isocenter was set to zero, ensuring that the incident photon field was correctly represented at the phantom surface. To simulate a full 180° breast CT acquisition, the phase-space file was virtually rotated around the phantom in 5° angular increments, rather than explicitly simulating each projection. This approach maintained the beam characteristics while significantly reducing computational cost and closely matching the experimental scanning protocol.

The simulations include detailed three-dimensional tracking of energy deposition within the breast phantoms, which were used for subsequent dosimetric analysis. The simulation setup and parameters were consistent with those described in Section 2.3.2. Voxel-specific absorbed dose values, accounting for material composition, were recorded in .3ddose files.

These data enabled accurate estimation of glandular dose under anatomically realistic conditions and formed the basis for subsequent analyses, including MGD estimation and DgN calculations, as described in the following sections.

### 2.5 DgN calculation

In this study, both the MGD and the incident $K_{air}$ were derived entirely from MC simulations, without reliance on physical measurements or empirical calibration factors. The DgN was then calculated as the ratio of MGD to incident $K_{air}$:

$$\mathrm{DgN} = \frac{\mathrm{MGD}}{\mathrm{K_{air}}} \qquad \text{(Equation 4)}$$

#### 2.5.1 $K_{air}$ estimation

$K_{air}$ was estimated using a direct energy-deposition scoring approach within the EGSnrc framework. To replicate the reference free-in-air geometry used in experimental dosimetry, the breast phantom was excluded from the simulation. Instead, a cubic air-scoring voxel with dimensions $3 \times 5 \times 1.5\ mm^3$ (volume = $22.5\ mm^3$) was placed at the simulation isocenter. This voxel corresponded to the sensitive volume of the Farmer chamber used in our experiments.

The air voxel was sufficiently small to satisfy charged-particle equilibrium (CPE) and to approximate a point-like measurement at the beam isocenter, while minimizing dose averaging across spatial beam gradients. Under conditions of CPE, which were satisfied for small air volumes irradiated by low-energy photons, the absorbed dose in the air voxel ($D_{air}$) is equivalent to the collision kerma. This equivalence has been validated in MC simulations [38,67], making voxel-based dose scoring a reliable method for estimating the incident air kerma. The absorbed dose in the air voxel was calculated as:

$$\mathrm{K_{air}} \approx \mathrm{D_{air}} = \frac{\sum_{i=1}^{N} E_i}{m} \qquad \text{(Equation 5)}$$

where $E_i$ is the energy deposited in the voxel during the i-th interaction, $N$ is the total number of energy-deposition events, and $m$ is the mass of the air voxel. The voxel mass was determined as $m = \rho_{air} V$, using an air density of $\rho_{air} = 1.204 \times 10^{-3}\ g\,cm^{-3}$ and a voxel volume of $V = 22.5\ mm^3$, yielding $m = 0.0525\ mg$. Each $K_{air}$ simulation employed $5 \times 10^9$ photon histories, resulting in statistical uncertainties below 1%.

#### 2.5.2 MGD calculation

For each breast specimen, the MGD was computed as the total energy deposited in glandular tissue divided by the total glandular mass. The three-dimensional dose distributions generated as described in Section 2.4 and stored in .3ddose format file were used to extract voxel-wise energy deposition data. For a given voxel i, the total deposited energy was:

$$E_i = \sum_{j \in i} E_{dep,j} \qquad \text{(Equation 6)}$$

where $j \in i$ denoted all photon and secondary particle interactions occurring within voxel i and $E_{dep,j}$ was the energy deposited by the j-th interaction. The total energy deposited in glandular tissue was then:

$$E_{gland} = \sum_{i \in gland} E_i \qquad \text{(Equation 7)}$$

The MGD was calculated as:

$$MGD = \frac{\sum_{i \in gland} E_i}{\sum_{i \in gland} M_i} \qquad \text{(Equation 8)}$$

where $M_i$ is the mass of glandular voxel i. By employing the same number of photon histories ($10^9$) as used for the $K_{air}$ simulations, MGD was estimated with similarly low statistical uncertainty (< 1%), ensuring the robustness of the derived $D_gN$ coefficients.

**2.6 Dose visualization and voxel-based analysis**

Spatial dose visualization and voxel-based analysis were performed using the Computational Environment for Radiotherapy Research (CERR) [68], a MATLAB-based toolkit. The reconstructed CT volumes and corresponding .3ddose files were imported into CERR, enabling spatially registered analysis of anatomy and dose.

Coronal dose color scale overlay was generated to visualize dose distributions within the breast phantoms. This facilitated qualitative assessment of dose uniformity, glandular dose localization, and spatial dose gradients, supporting the quantitative analyses described in the subsequent sections.

**3. Results**

**3.1 Verification of MC imaging beam modelling**

To validate the MC model of the BCT imaging system, the PDD curves obtained from the experimental water tank measurements were compared with the corresponding PDD curves calculated using the EGSnrc/DOSXYZnrc simulation framework. PDDs were evaluated at multiple positions across the beam, including central and off-axis locations, to confirm spatial consistency at a given energy. Agreement between PDDs exceeded 99% at all positions. Therefore, for clarity, only the central-axis PDD is presented here as a representative validation metric. Figure 1 presents the comparison between the measured and simulated PDD curves at 32 keV, which represented the typical beam energy used for imaging experiments. Both curves were normalized to their respective maximum dose values to facilitate direct comparison. Overall, the PDD curves demonstrated excellent agreement throughout the entire depth range, with minor discrepancies near 3–4 cm depths. Quantitative agreement between the measured and simulated PDD curves was assessed using gamma index analysis with criteria of 2 mm DTA and 2% dose difference. Although PDD is a 1D metric, gamma analysis along the central axis is still a relevant quantitative comparison method, as discussed in Ref. [66]. The resulting gamma index distribution is shown in Figure 1. 21 of 22 points (96%) satisfied the gamma criterion ($\gamma < 1$), demonstrating strong overall agreement between the measured and simulated dose distributions. Similar agreement was observed for the other beam energies, with 97%, 96%, and 96% of points passing the gamma criterion for 30, 35, and 37 keV, respectively (graphs not shown).

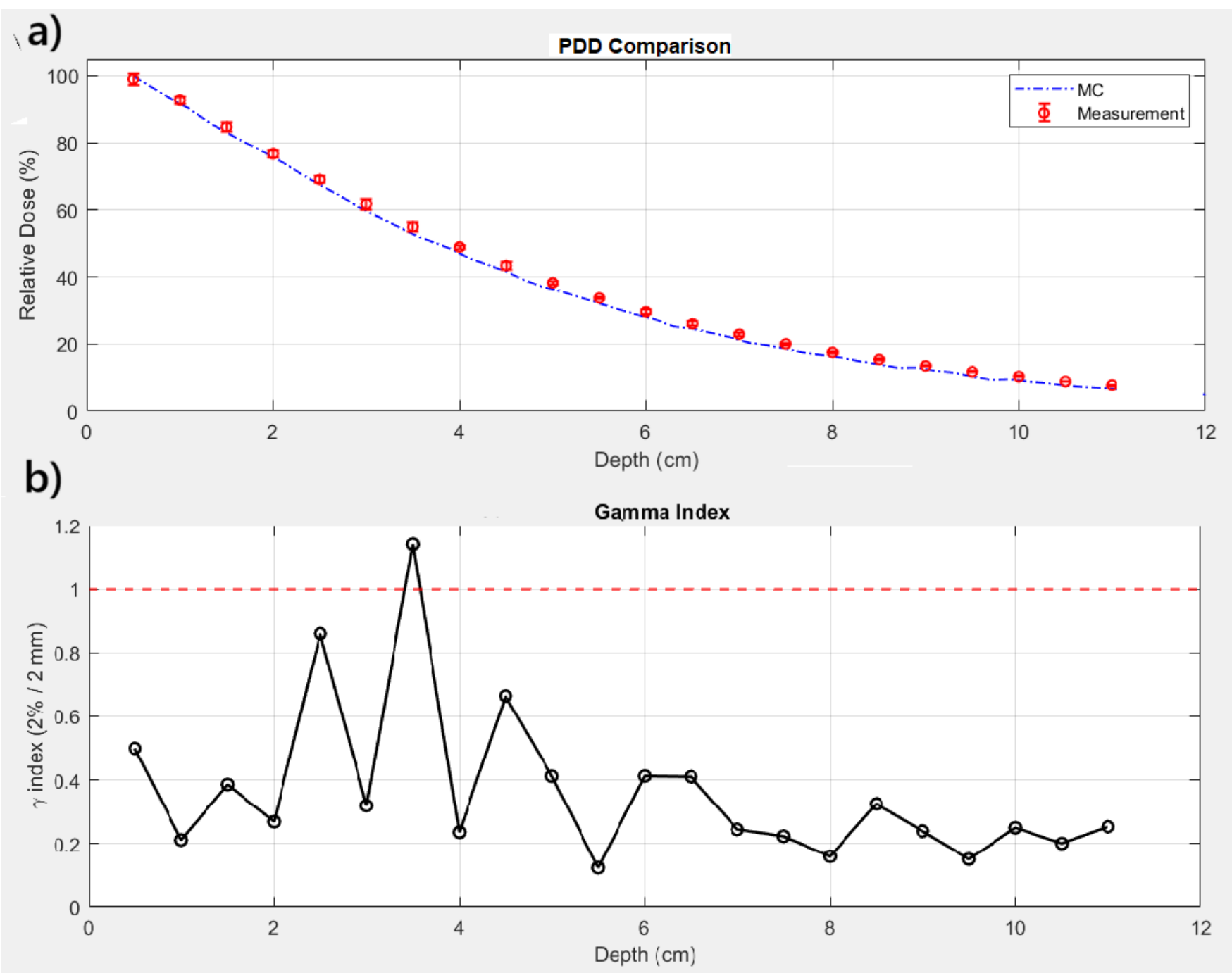

*Figure 1. (a) Comparison of measured (red dots) and MC-calculated model (purple line) PDD curves in the water tank at 32 keV. (b) Gamma index (blue line) between measured and calculated curves, demonstrating agreement of evaluated points.*

#### 3.2. MGD evaluation

Figure 2a presents MGD values as a function of photon energy for heterogeneous breast phantoms with a 1 mm skin layer and a constant breast diameter of 11.4 cm. The analysis was performed for Samples 1 and 2, both with a height of 6.5 cm and glandular fractions of 10% and 25%, respectively. The MGD was normalized to the number of primary photons in the MC simulations. Photon energy varied from 28 to 38 keV in 2 keV increments while maintaining a constant photon fluence. As shown in Figure 2a, Sample 2, with higher glandularity (25%), exhibited lower MGD values than Sample 1, which had lower glandularity (10%). This indicated that, for a constant breast height and constant incident air kerma, increasing glandularity led to a reduction in MGD.

Figure 2b shows MGD as a function of photon energy for Samples 3–5 using realistic heterogeneous breast models with a 1 mm skin layer. The heights of Samples 3–5 were 5.0, 7.5, and 9.0 cm, respectively. Larger breast height led to higher MGD values. Considering the trend observed in Figure 2a, where lower glandularity corresponded to higher MGD, the overall decrease in MGD from Samples 5 to 3 can be attributed to the combined effects of decreasing breast height and increasing glandularity, both of which act to reduce MGD.

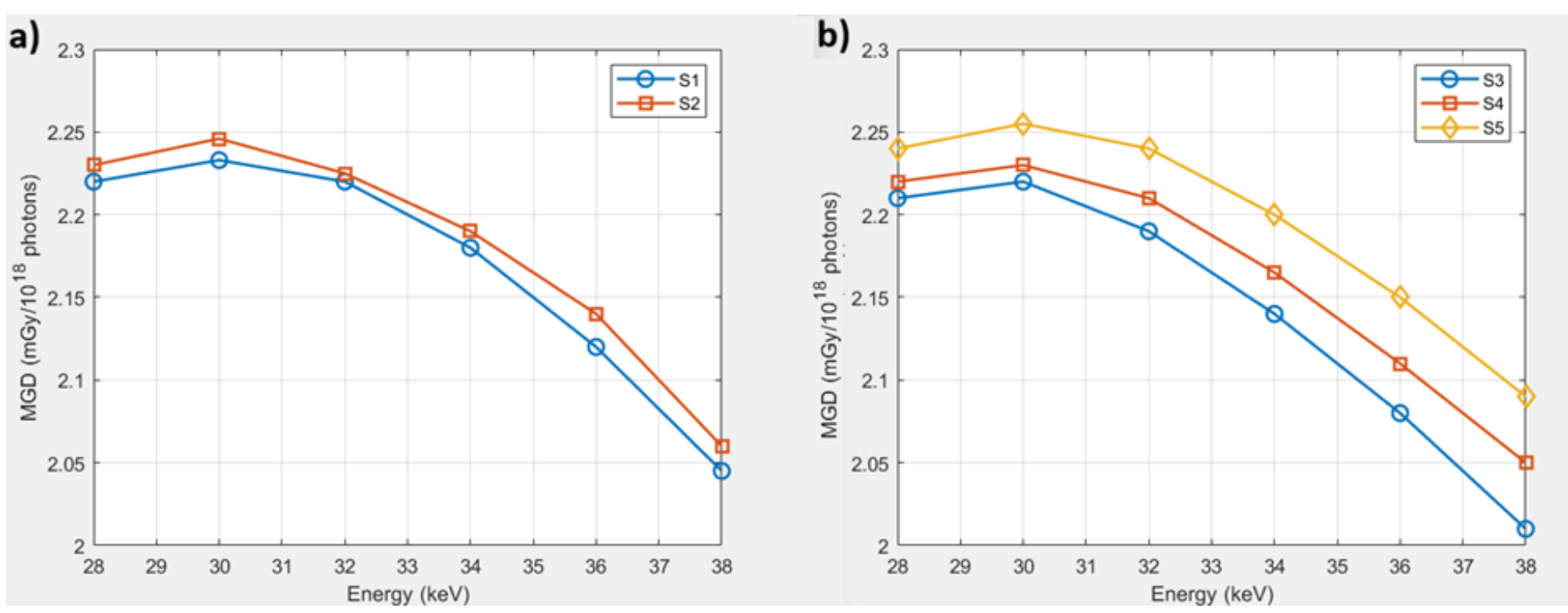

***Figure 2. MGD values as a function of photon energy for heterogeneous breast phantoms. (a) Samples 1 and 2 (orange and blue, respectively), with identical breast height of 6.5 cm but different glandular fractions of 10% and 25%, respectively. (b) Samples 3–5 (blue, red, and yellow, respectively), with breast heights of 5 cm, 7 cm, and 9 cm, and corresponding glandular fractions of 40%, 30%, and 25%, respectively.***

#### 3.3 DgN evaluation

#### 3.3.1. Effect of glandularity and breast height

Figure 3 shows the DgN coefficients for Samples 3–5 over the primary photon energy range. Results are presented for heterogeneous and material-homogeneous phantoms with 1–3 mm skin thickness. For all phantoms and skin thicknesses,

DgN (MGD/K) increased with photon energy over the range of 28–38 keV. At all energies, DgN increased with increasing breast height and decreasing glandularity. This behavior was consistent across the investigated energy range and indicated that both increasing glandularity and decreasing breast height reduce DgN.

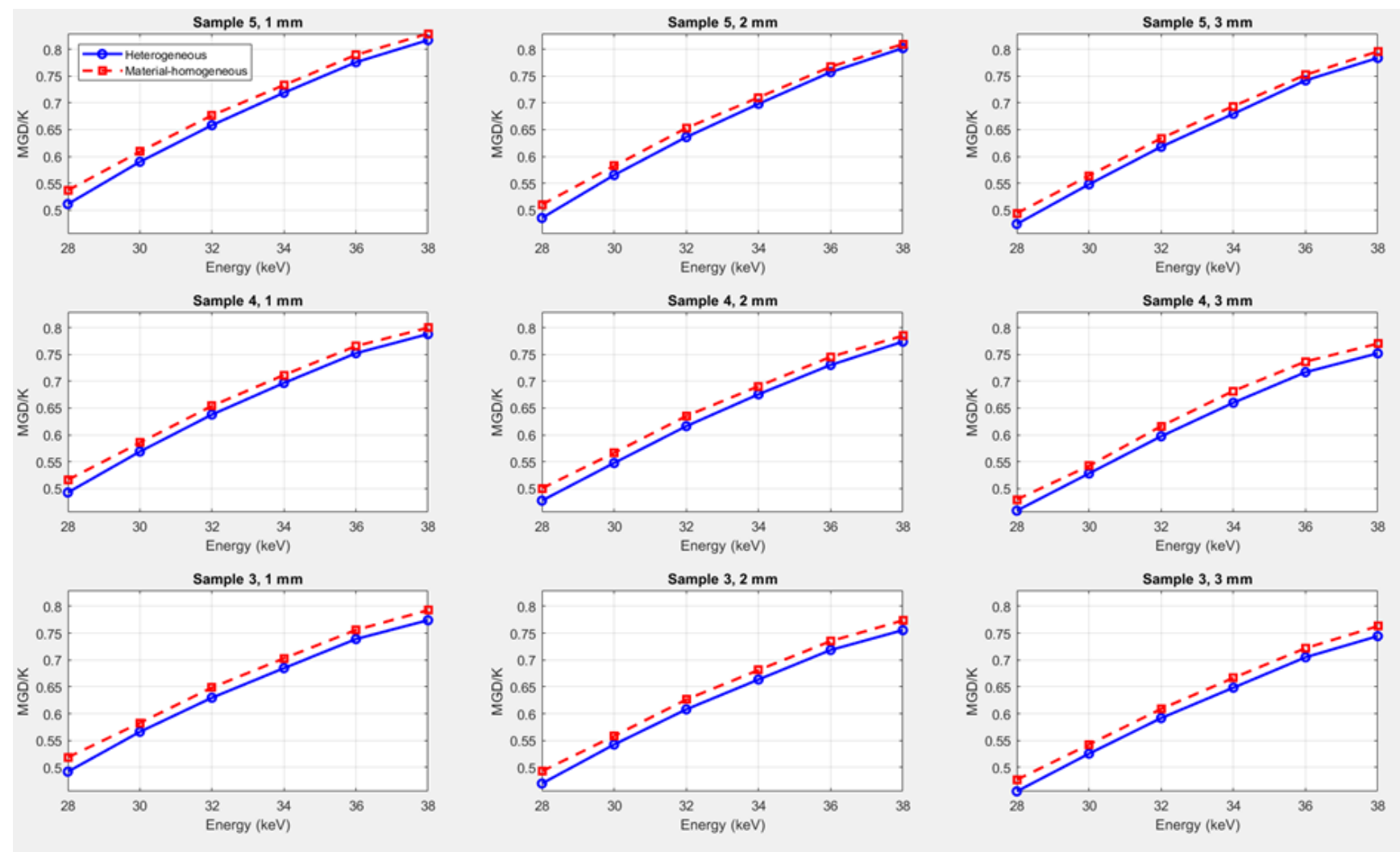

***Figure 3. DgN (MGD/K) as a function of photon energy for Samples 3–5 (bottom to top curves), comparing heterogeneous (blue) and material-homogeneous (red) breast phantom results. Results are shown for skin thicknesses of 1, 2, and 3 mm (left to right panels).***

To further quantify the influence of breast height and glandularity observed in Figure 3, Sample 4 and Sample 5 were directly compared (Figure 4) by calculating the percentage difference in their DgN coefficients for heterogeneous phantoms with identical skin thicknesses (1–3 mm). The percentage difference was computed for each photon energy by comparing the corresponding DgN values of these two phantoms. The results showed that the difference increased progressively with photon energy, rising from approximately 1% at 28 keV to nearly 5% at 38 keV. This behavior was observed consistently across all skin thickness configurations. The effect was more pronounced for thinner skin layers, where reduced skin attenuation increased the relative contribution of the internal breast tissue composition to the total MGD.

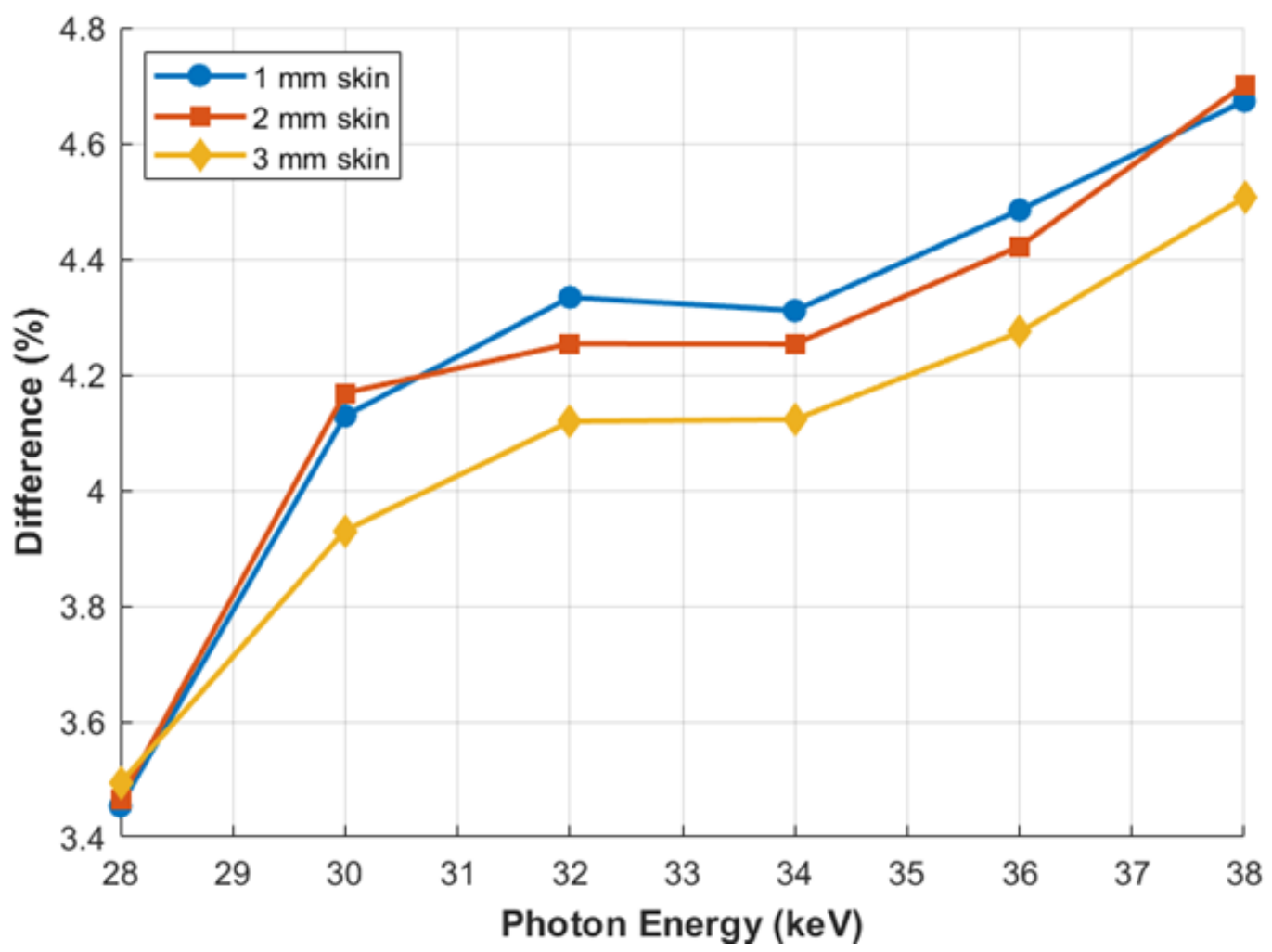

***Figure 4. Percentage difference in DgN coefficients between Samples 5 and 4 as a function of photon energy for heterogeneous breast phantoms with skin thicknesses of 1, 2, and 3 mm (blue, red, and yellow, respectively).***

**3.3.2. Effect of phantom type**

In Figure 3, the DgN coefficients for the material-homogeneous and patient-specific heterogeneous phantoms were compared, illustrating the effect of phantom composition on the calculated DgN values. The results indicated that the material-homogeneous model led to an average MGD overestimation of approximately 3%. The discrepancy was more pronounced at lower photon energies, particularly for Sample 3 with 1 mm skin thickness at 28 keV, where the maximum observed difference between the two models reached 5%.

**3.3.3. Influence of skin thickness**

Figure 3 also demonstrates the influence of skin thickness (1, 2, and 3 mm) on the DgN coefficient for three representative breast samples across the investigated photon energy range. Increasing skin thickness resulted in a systematic reduction in DgN, with up to a 10% decrease observed when the skin layer increased from 1 mm to 3 mm. This trend was consistent across all photon energies and breast samples, confirming that the skin acts as an attenuating layer that reduces the dose reaching the underlying glandular tissue.

Figure 5 further presents the ratio of DgN values calculated for the heterogeneous breast model with 2 mm and 3 mm skin layers relative to a 1 mm skin layer for Sample 1. When skin thickness increased from 1 mm to 3 mm, DgN decreased by approximately 9.0% at 28 keV, 6.1% at 34 keV, and 3.6% at 38 keV. For a 2 mm skin layer, the corresponding reductions were about 5.0%, 3.5%, and 2.3% at these energies, respectively. Moreover, the percentage difference between the heterogeneous and material-homogeneous models remained nearly 4–5% at each photon energy as skin thickness increased (Figure 3). This suggested that the attenuation introduced by the skin layer was largely independent of the internal composition of the phantom.

Overall, these results confirmed that increasing skin thickness systematically reduces the glandular dose (for a fixed incident $K_{air}$), and that this attenuation effect becomes less pronounced at higher photon energies.

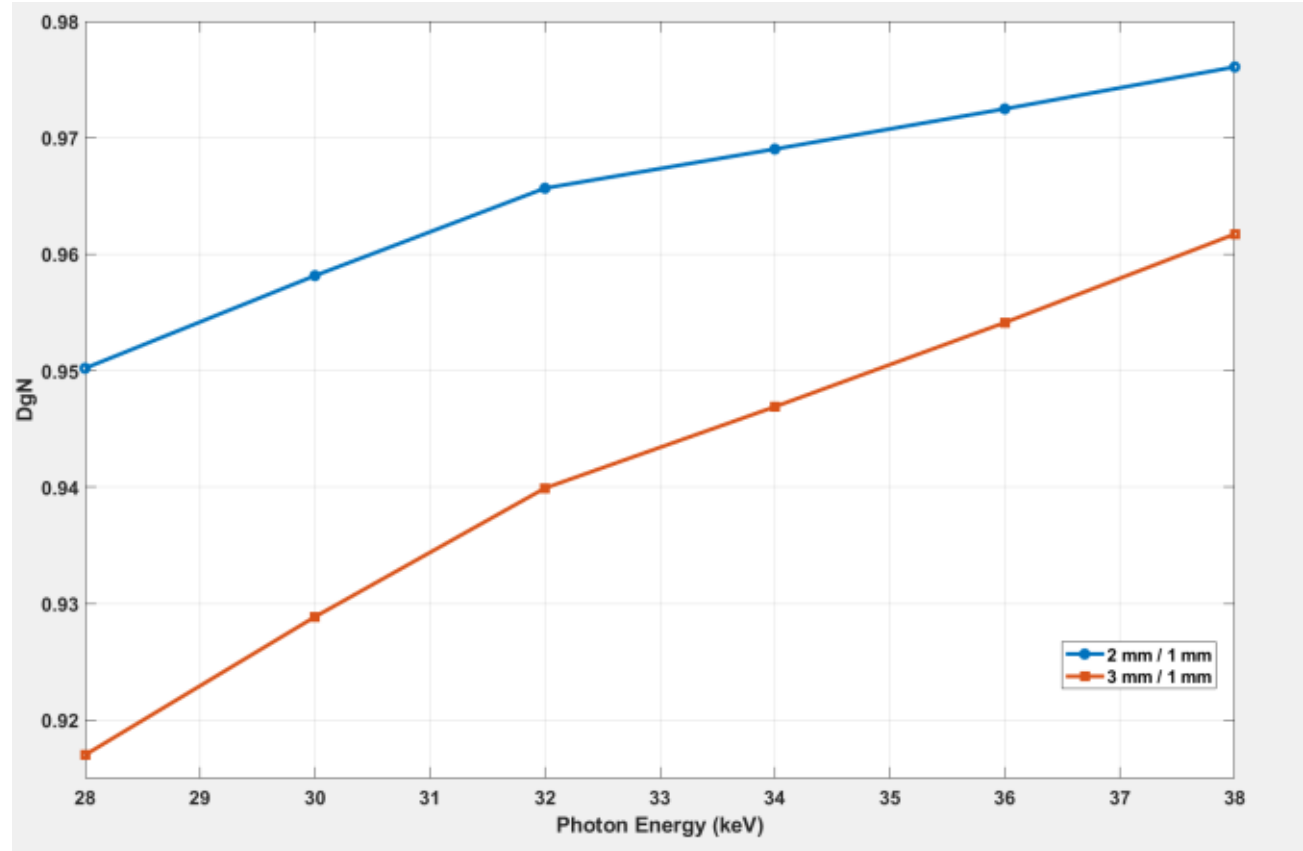

***Figure 5. Ratio of DgN coefficients for heterogeneous breast models with 2 mm and 3 mm skin thicknesses (blue and red, respectively) relative to the 1 mm skin configuration for Sample 1.***

**3.3.4 Homogeneous models vs. heterogeneous phantom**

The DgN coefficients obtained from the patient-specific heterogeneous breast phantom were compared with those derived from the corresponding homogeneous breast model for Sample 1, considering skin thicknesses of 1–3 mm (Figure 6a). Additionally, the skinless homogeneous model for Sample 1 was compared with the corresponding MATLAB-based homogeneous phantom (Figure 6b) to validate the implementation and consistency of the homogeneous modelling approach.

Quantitative analysis in Figure 6a shows that the homogeneous model systematically underestimated the DgN values compared to the patient-specific heterogeneous phantom. For the 1 mm skin layer, the DgN obtained from the homogeneous model was approximately 36% lower at 38 keV, while the difference decreased to about 25% at 28 keV. For the 3 mm skin layer, the underestimation was reduced to approximately 25% at 38 keV and about 17% at 28 keV. This reduction in discrepancy with increasing skin thickness indicated that the influence of internal tissue heterogeneity on DgN becomes less pronounced as the attenuating effect of the skin layer increases.

Comparison between the homogeneous model and the MATLAB-based homogeneous phantom demonstrated excellent agreement, with an average consistency of approximately 99% in DgN values. This strong agreement confirmed the reliability and correct implementation of the homogeneous MGD calculation framework.

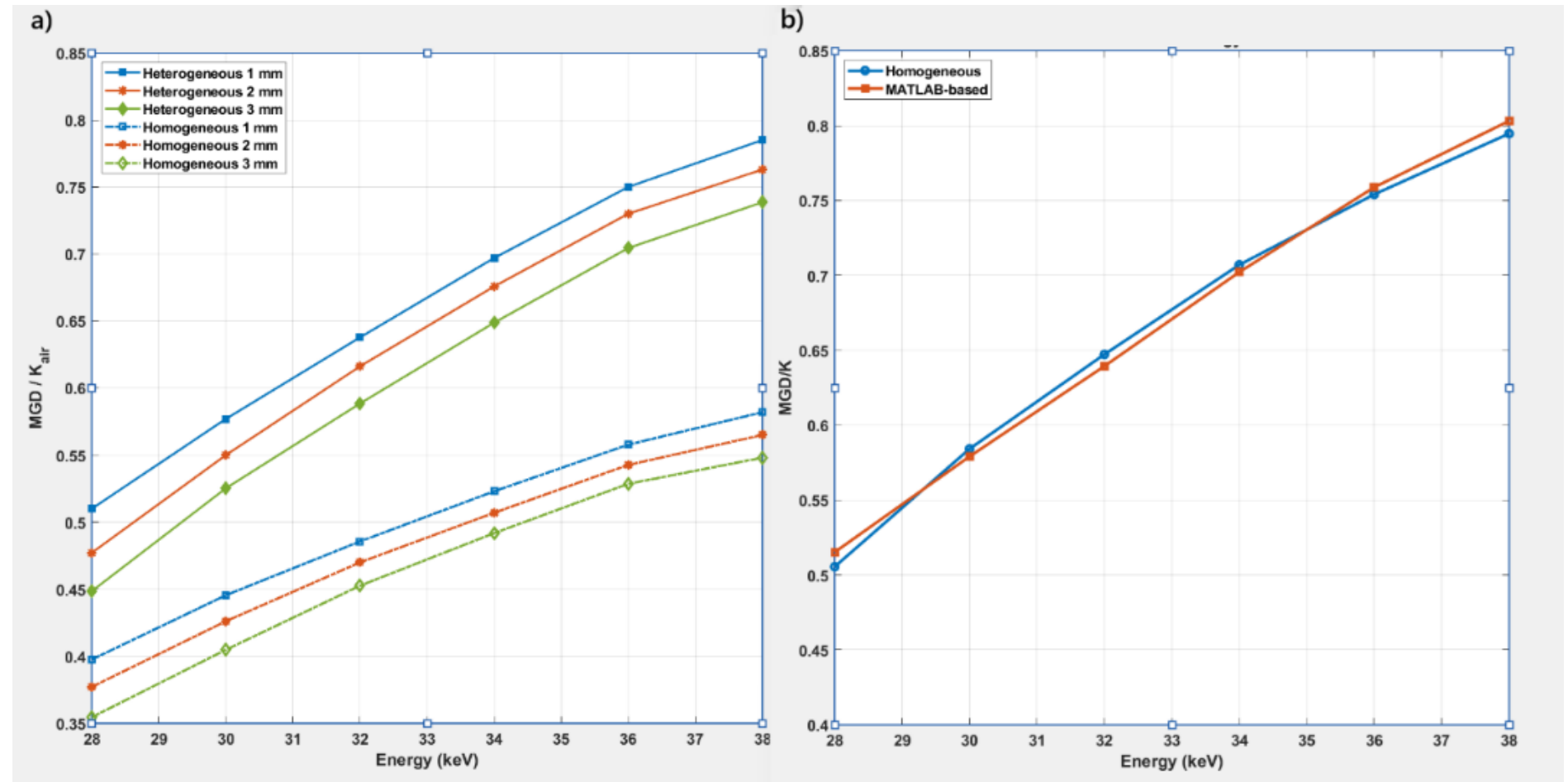

***Figure 6. (a) Comparison of DgN coefficients between the patient-specific (Sample 1) heterogeneous and homogeneous breast phantoms for skin thicknesses of 1, 2, and 3 mm (blue, red, and green, respectively). (b) Validation of the homogeneous model against the MATLAB-based homogeneous phantom (blue and red curves, respectively).***

In addition to the quantitative comparison of DgN values, spatial glandular dose distributions were evaluated for the 2 mm skin configuration at an incident $K_{air}$ of 1 mGy (Figure 7). Figure 7a shows a representative reconstructed slice of the breast sample data set, with glandular, adipose, and skin regions identified. The corresponding dose distributions for the heterogeneous and homogeneous phantoms are presented in Figures 7b and 7c, respectively. Both homogeneous and heterogeneous phantoms exhibited similar dose distribution patterns. In all cases, higher glandular dose deposition was observed near the peripheral regions of the breast, with progressively lower doses toward the central regions. The homogeneous model and the MATLAB-based homogeneous phantom showed nearly identical spatial dose patterns (Figures 7c and 7d), further supporting the consistency of the homogeneous models. Furthermore, the dose maps clearly demonstrated that the absorbed dose in glandular tissue was higher than in adipose tissue, as indicated by the red regions corresponding to glandular voxels. This higher dose deposition in glandular regions was consistently observed in all phantom types.

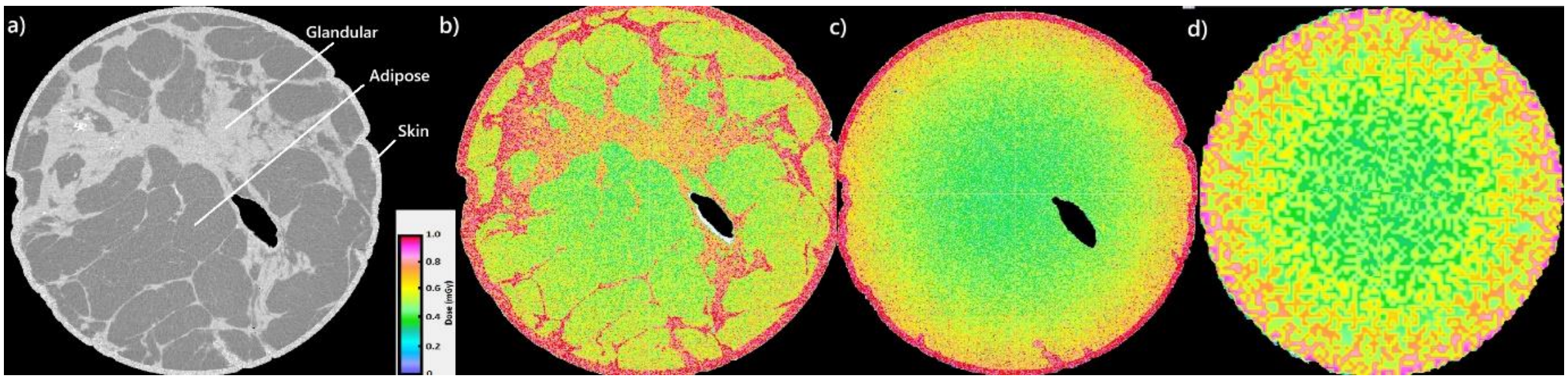

**Figure 7. (a) Representative reconstructed slice of the breast sample with glandular, adipose, and skin regions indicated. (b) Calculated dose distribution for the patient-specific heterogeneous phantom. (c) Dose distribution for the corresponding homogeneous breast model. In (b) and (c), the colour scale indicates a maximum dose of 1 mGy in red and a minimum dose in purple. (d) Dose distribution obtained from the MATLAB-based homogeneous phantom.**

**4. Discussion**

This study demonstrates that patient-specific breast phantoms, combined with a validated and system-tailored MC beam model, provide an accurate framework for MGD estimation. By explicitly modelling realistic breast anatomy, tissue composition, and imaging geometry, this approach establishes a gold-standard reference against which simplified breast models can be rigorously evaluated. In this work, the validated patient-specific framework enables a systematic assessment of homogeneous breast models and can be extended to evaluate improved population-based models, such as Boone *et al.* (2017), Hernandez *et al.* (2015), and Tucciariello et al. (2021) which incorporate more realistic breast anatomical characteristics into phantoms based on breast CT data [69,70,34].

Unlike conventional approaches that rely solely on MGD, which represents a spatial average over the entire glandular volume, the present work resolves the full three-dimensional dose distribution within realistic heterogeneous breast models. Because X-ray attenuation and tissue heterogeneity produce inherently non-uniform energy deposition, certain glandular subregions can receive doses substantially higher than the reported MGD, information that is obscured by averaging. Since radiation-induced cancer risk is also governed by localized cellular dose rather than global mean values [71], voxel-level dosimetry provides a more physically and biologically relevant assessment of risk. This constitutes a significant methodological advantage over simplified models that depend exclusively on MGD.

The agreement between EGSnrc-simulated and experimentally measured PDD confirms that the MC modelling accurately reproduces the physical beam characteristics required for reliable MGD estimation. The minor deviations observed between MC-calculated and experimentally measured PDD can be attributed to the statistical nature of MC simulations and inherent uncertainties in water tank measurements. Considering these uncertainties, the achieved 96% gamma agreement confirms the reliability of the phase-space-based MC beam model for subsequent dose simulations. With this validation established, the analysis of MGD behavior under varying breast parameters and X-ray energy range can be interpreted with confidence. To further support the validity of the MC modelling, the DgN values obtained in this study were compared with the most comparable case reported in the literature [45], showing overall agreement within 4%. Some discrepancies are expected due to differences in imaging conditions, such as the larger breast diameter used in this study and the use of a synchrotron-based parallel beam rather than a cone-beam geometry. Despite these differences in imaging modality and phantom construction, the consistency with previously published DgN values further supports the accuracy of the MC-based dose calculations.

The dependence of MGD on both glandularity and breast height is consistent with previously reported findings (Refs. 45, 36, and 67). For constant glandularity, increasing breast height results in higher MGD (for a fixed incident air kerma) due to the larger irradiated tissue volume and increased total energy deposition per unit glandular mass. This behaviour is physically intuitive and aligns with established dosimetric principles.

With respect to glandularity, glandular tissue exhibits higher density and attenuation compared to adipose tissue. Since MGD is defined as the mean absorbed energy to the glandular tissue divided by total glandular mass, increasing glandularity increases glandular mass approximately linearly, while total absorbed energy increases more moderately. This imbalance results in a net decrease in MGD with increasing glandularity. This effect is further amplified in rotational breast CT geometry, where irradiation occurs from all directions. Increased glandular content enhances self-shielding: glandular tissue attenuates incoming photons before they reach deeper regions. Consequently, photon fluence within central glandular regions is reduced, limiting absorbed energy in these regions. The combined effects of increased glandular mass and reduced photon fluence result in a net decrease in MGD as glandularity increases (for a fixed incident air kerma).

When breast height and glandularity vary simultaneously, as observed in Samples 3–5, the resulting MGD reflects the combined influence of these factors. The interplay of these mechanisms explains the behavior observed in Figure 2 and Figure 3.

To evaluate simplified dosimetric models, DgN coefficients derived from patient-specific heterogeneous phantoms were compared with those from corresponding homogeneous phantoms. The homogeneous phantom type systematically underestimated DgN relative to the heterogeneous reference. This underestimation is clinically significant, as it may lead to systematic bias in MGD estimation. Specifically, an underestimated DgN coefficient directly results in a proportionally lower calculated MGD for a given $K_{air}$, based on Equation (1), which can in turn lead to the delivery of a higher actual patient dose than intended when exposure settings are determined based on the underestimated value. The energy-dependent variations further emphasize the limitations of homogeneous representations in capturing patient-specific dose characteristics. Neglecting anatomical heterogeneity, particularly the spatial distribution of glandular tissue, introduces measurable bias.

A comparison between material-homogeneous and fully heterogeneous phantoms further clarifies the origin of the observed discrepancies. In the material-homogeneous model, each segmented tissue type was assigned a single mean density, thereby removing voxel-by-voxel density variations. Under these conditions, the difference relative to the corresponding heterogeneous phantom was limited to a maximum of 5%. In contrast, comparisons between homogeneous and heterogeneous models showed differences of up to 36%. Because removing voxel-scale density variations produced only minor changes, whereas removing the anatomical distribution of glandular tissue produced large deviations, the spatial arrangement of glandular tissue emerged as the primary determinant of glandular dose. These findings demonstrate that accurate MGD estimation depends predominantly on anatomical structure rather than average tissue composition alone. The generation and investigation of the material-homogeneous phantom were essential to establishing this conclusion.

The excellent agreement between the MATLAB-based phantom and the voxelized homogeneous phantom confirms the numerical stability and implementation accuracy of the model. This validation establishes the MATLAB-based homogeneous approach as an independent reliable tool for simplified dose estimation.

The results presented in this work demonstrate that the location and concentration of glandular regions significantly influence MGD. Building on these findings, future work will focus on more precise characterization of breast properties, particularly the spatial concentration and distribution of glandular tissue, as well as on developing correction factors that relate simplified MATLAB-based homogeneous MGD estimates to anatomically realistic heterogeneous phantoms. The MATLAB-based platform provides a controlled and computationally efficient environment for establishing such relationships without requiring patient-derived phantoms.

This work aims to demonstrate the feasibility of a patient-specific MC dosimetry framework rather than to establish population-level dose statistics. An important next step is the development of a comprehensive library of realistic breast phantoms derived from volunteer imaging data, covering a wide range of breast sizes, glandularities, and anatomical variations. Expanding the dataset to represent diverse breast morphologies would enable the creation of population-representative lookup tables for MGD conversion factors. Such resources could facilitate clinically practical patient-specific dose estimation by linking measurable breast parameters (e.g., size, density, and distribution of glandular) to validated conversion coefficients derived from anatomically realistic models. Ultimately, this approach could improve the accuracy and broader applicability of MGD estimation in breast CT and other dedicated breast imaging modalities.

**5. Conclusion**

By integrating patient-specific anatomical phantoms with a validated beam model, the proposed MC framework enables calculation-based estimation of MGD under realistic synchrotron-based imaging conditions. The results demonstrate that breast geometry, glandular distribution, and skin thickness significantly influence DgN and MGD, while simplified homogeneous models introduce systematic bias, particularly when anatomical heterogeneity is neglected. These findings underscore the importance of anatomically realistic modelling for reliable dosimetry in advanced breast imaging systems. The proposed approach provides a robust reference platform for patient-specific dose assessment and supports the safe clinical synchrotron-based breast CT. Notably, this work demonstrates the capability of the EGSnrc MC code, which has not previously been widely applied in this context, to model and quantify dose in such complex imaging scenarios.

**6. Acknowledgements**

The authors sincerely thank all participants who generously donated their tissue for this study, without whom this research would not have been possible. This work was supported by the National Health and Medical Research Council (Grant No. 2021/GNT2011204). This research was undertaken on Imaging and Medical beamline at the Australian Synchrotron, part of ANSTO we are grateful for the access to the facility provision of beamtime and the dedication of support staff to make this research possible.